\documentclass[aps,prl,reprint,superscriptaddress]{revtex4-1}
\usepackage{graphicx}

\begin{document}

\title{Silver catalyzed Fluorouracil degradation; a promising new role for graphene}

\author{Federico~Mazzola}

\affiliation{Department of Physics, Norwegian University of Science and Technology (NTNU), N-7491 Trondheim, Norway }
\author{Thuat~Trinh}
\affiliation{Department of Chemistry, Norwegian University of Science and Technology (NTNU), N-7491 Trondheim, Norway }

\author{Simon Cooil}
\affiliation{Department of Physics, Aberystwyth University, Aberystwyth SY23 3BZ, UK}

\author{Elise Ramleth \O stli}
\affiliation{Department of Materials Science and Engineering
Norwegian University of Science and Technology (NTNU), N-7491 Trondheim, Norway}

\author{Kristin H\o ydalsvik}
\affiliation{Department of Chemistry and Molecular Biology, University of Gothenburg, SE-412 96 Gothenburg, Sweden}

\author{Eirik Torbj\o rn Bakken Skj\o nsfjell}
\affiliation{Department of Physics, Norwegian University of Science and Technology (NTNU), N-7491 Trondheim, Norway }



\author{Signe Kjelstrup}
\affiliation{Department of Chemistry, Norwegian University of Science and Technology (NTNU), N-7491 Trondheim, Norway }

\author{Alexei Preobrajenski}
\affiliation{MAX IV, Lund University, Box 118, 22100 Lund, Sweden}

\author{Attilio A. Cafolla}
\affiliation{School of Physical Sciences, Dublin City University, Dublin 9, Ireland}

\author{D.~Andrew Evans}
\affiliation{Department of Physics, Aberystwyth University, Aberystwyth SY23 3BZ, UK}

\author{Dag W. Breiby}
\affiliation{Department of Physics, Norwegian University of Science and Technology (NTNU), N-7491 Trondheim, Norway }

\author{Justin~W.~Wells}
\email[]{quantum.wells@gmail.com}
\affiliation{Department of Physics, Norwegian University of Science and Technology (NTNU), N-7491 Trondheim, Norway }

\date{\today}

\begin{abstract}
Chemotherapy treatment usually involves the delivery of fluorouracil (5-Fu) together with other drugs through central venous catheters. Catheters and their connectors are increasingly coated (or impregnated) with silver or argentic alloys/compounds.  Complications such as broken catheters are common, leading to additional suffering for patients and increased medical costs. Here, we uncover a likely cause of such failure through a study of the surface chemistry relevant to chemotherapy drug delivery, i.e. between 5-Fu and silver.  We show that silver catalytically decomposes 5-Fu, releasing HF as a product. This reaction compromises the efficacy of the treatment, and at the same time, releases HF which is damaging to both patient and catheter.  Our study not only reveals an important reaction which has so far been overlooked, but additionally allows us to propose that graphene coatings inhibit such a reaction and offer superior performance for cancer treatment applications.
\end{abstract}

 \maketitle

\begin{figure}[b!]
\centering
\includegraphics [width=\columnwidth]{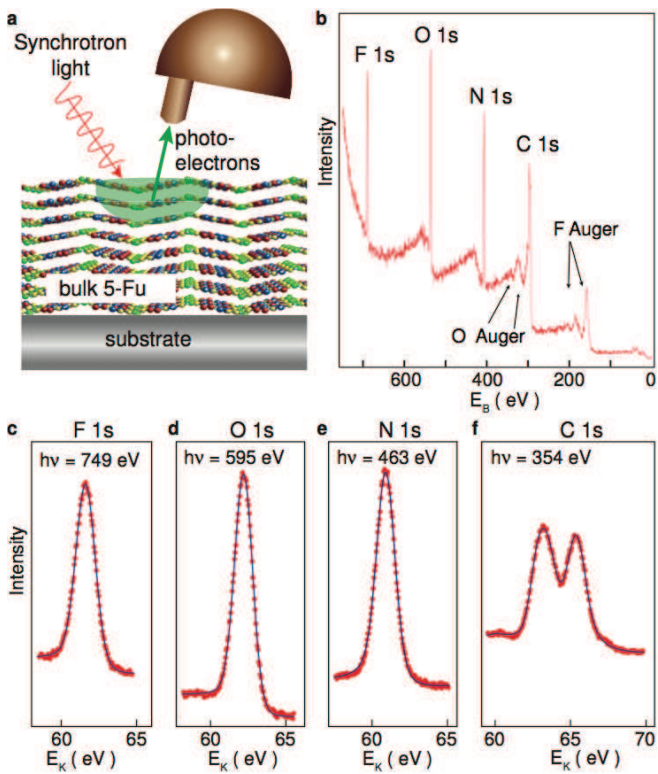}
\caption{\textbf{Photoemission spectroscopy of a bulk-like 5-Fu sample:} \textbf{a,} Schematic of the photoemission process and the depth sensitivity of the experiment. \textbf{b,} An overview measurement of the bulk-like 5-Fu sample (photon energy $h\nu$=800~eV, temperature -170~$^{\circ}$C). \textbf{c-f,} Detail of the F1s, O1s, N1s and C1s regions respectively, measured with lower kinetic energy and higher resolution. The raw experimental data (red) and a fit to the data (blue). Details of the fit and of the quantitative analysis are reported in the supplementary material.} \label{fig1}
\end{figure}


 
5-Fluorouracil (5-Fu) is one of the most commonly used drugs in chemotherapy treatments \cite{Lu:1993,Sistigu:2014,Longley:2003,Levine:1998}.  5-Fu acts to  inhibit  thymidine, a nucleoside required for DNA replication and necessary for cancer cells to reproduce \cite{Longley:2003}. The administration of 5-Fu into the human body is generally accomplished via a central venous catheter. Catheters are made of polymeric materials such as polyurethane and silicone \cite{OLeary:2011} and they are prone to degradation when in contact with bodily fluids or aggressive drugs such as 5-Fu  \cite{Fakler:2004, Ardalan:1995}. Degradation can be  reduced by applying protective coatings onto the internal and/or external surfaces of the catheters \cite{OGrady:2011, Masci:2003}. Among the materials used for coating purposes, noble metals (i.e.\ silver and argentic alloys  \cite{Timsit:2011,Armentano:2014, Suska:2010, Harter:2002}) are particularly common because of their low reactivity and antimicrobial properties \cite{Maki:1997, Singh:2008, Guggenbichler:1999, Feng:2000}.  Despite the efforts made to improve the quality of the treatments of cancer and the life expectancy of the patients, investigation of a possible chemical reaction between coating materials and the chemotherapeutic drug itself is lacking (although some studies of the electronic interaction exist, see for example Ref.\ \onlinecite{Pavel:2006}). In this work, we study the degradation of 5-Fu by silver surfaces, we report that HF is formed as a product, and we demonstrate graphene coatings to be an inert alternative.

\section*{Results}

We initially present an X-ray photoemission spectroscopy (XPS) investigation of `bulk-like' 5-Fu. Although XPS is a powerful technique for studying organic molecules  (including uracil \cite{Papageorgiou:2012}), to the best of our knowledge no XPS studies of 5-Fu exist. Since the depth sensitivity in an XPS experiment is limited by the mean-free-path of the photoemitted electrons (typically $\approx$1~nm), a 4~nm thick film of 5-Fu is sufficiently thick to be considered bulk-like. Indeed, our XPS study reveals that only the molecules, and not the underlying substrate, contribute to the signal (the XPS technique and depth sensitivity are represented in Fig.\ \ref{fig1}a, and an overview XPS measurement in Fig.\ \ref{fig1}b).   In our investigation of this bulk-like film, we also measured the 1s core level spectra for nitrogen, fluorine, oxygen and carbon at low kinetic energy, and these results are summarized in Fig.\ \ref{fig1}c-f.

Quantitative analysis of the XPS data allows us to confirm the stoichiometry of the molecular film. 5-Fu is composed of C, F, N and O in a ratio $4:1:2:2$, and this is in good agreement with the experimentally determined abundance $4.6:1:2.2:1.8$ (measured relative to F, which is defined as unity).  The multiple contributions to the C1s core level indicate that carbon is present in multiple bonding configurations with dissimilar binding energies. Furthermore, we carried out a similar XPS study on a bulk-like film prepared from powder (i.e. not thermally evaporated in vacuum) to show that 5-Fu is not damaged by evaporation (see supplementary material).

\begin{figure*}
\centering
\includegraphics [width=1\textwidth]{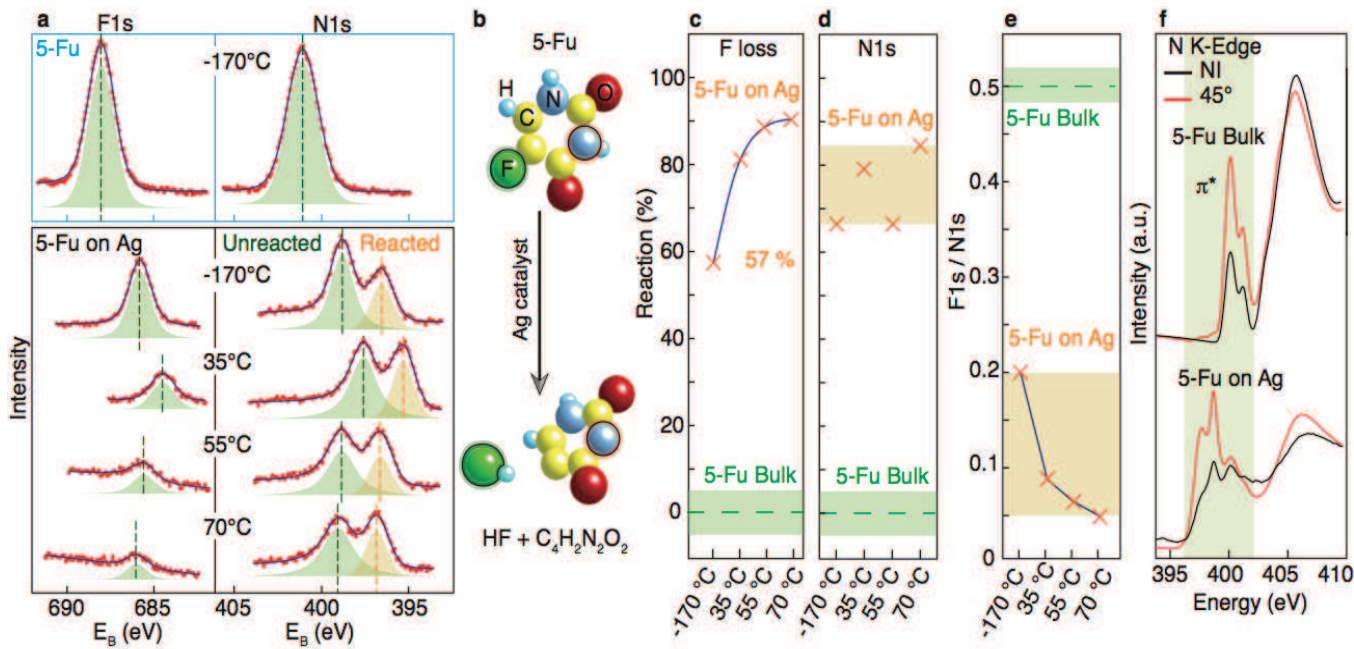}
\caption{\textbf{The reaction of 5-Fu with an Ag substrate:} \textbf{a,} XPS measurements  of the N1s and F1s core levels of bulk 5-Fu (upper panel) and 5-Fu on Ag (lower panel). The green (orange) color indicates unreacted (reacted) components.  Raw data (red) and fits (blue) are shown. The reacted component of the N1s spectra appears at lower binding energy ($\approx$2.2~eV with respect to the unreacted component). Both F1s and N1s show a shift (due to a charge transfer induced by doping) indicated by the dashed lines. The intensities of the core levels have been normalized to the synchrotron beam current and acquisition time. The photo-ionization cross section has been considered for the quantitative analysis (see supplementary material). \textbf{b,} Schematic of the proposed reaction. The atoms involved in the reaction have been highlighted.  \textbf{c,} The proportion of F lost from the surface.    \textbf{d,}  The proportion of 5-Fu molecules containing a reacted N atom. \textbf{e,}  The F1s/N1s ratio as function of temperature. \textbf{f,}  Nitrogen XAS measurements at two different incidence angles for 5-Fu on Ag revealing a change in the $\pi$-bonding as compared to a bulk-like sample. } \label{fig2}
\end{figure*}

A thin film ($\approx$0.7~nm) film of 5-Fu was also prepared on an Ag substrate. XPS measurements (Fig.\ \ref{fig2}a) reveal that a reaction takes place at body temperature ($\approx$35~$^{\circ}$C). Measurement were also performed at lower temperatures in order to reduce the thermal energy available (i.e. to facilitate observation of the intermediate steps of the reaction and their approximate energy budget).

Our study reveals that already at -170~$^{\circ}$C the N1s core level undergoes a significant change from its bulk form, reflecting a change in the N bonding configuration. A new `reacted' N1s core level component is observed at larger binding energy, shifted 2.2~eV from the unreacted component. We propose that the reaction involves breaking one of the N-H bonds with a subsequent H loss from the molecule.  Assuming that only one N-atom per molecule is able to react, quantitative analysis of the N1s components indicates that $\approx$80$\%$ of the molecules are in the reacted form (Fig.\ \ref{fig2}d). The reaction is already favourable at the lowest temperature studied (-170~$^{\circ}$C), and thus increasing the temperature is not seen to play a significant role, as the reaction is completed far below body temperature.

Analysis of the F1s peak also reveals that a reaction is taking place -- seen as a reduction of the F1s intensity, indicating that fluorine is lost into the vacuum. Here the reaction shows a strong temperature dependence (Fig.\ \ref{fig2}c); at low temperature, the loss is $\approx$57$\%$ (relative to the bulk-like film), and as the temperature is increased to body temperature and above, the loss is almost complete ($\approx$90$\%$). The same behavior can also be observed by extracting the F1s/N1s intensity ratio (=0.5 for bulk 5-Fu) which is reduced to $\approx$0 as the temperature is increased (Fig.\ \ref{fig2}e).


Summarizing, we observe that 5-Fu in contact with Ag gives rise to reactions which modify the bonding of an N-atom in the 5-Fu molecule, and involves a massive F loss under vacuum conditions. We postulate a possible reaction which is consistent with these observations (Fig.\ \ref{fig2}b). Whilst the modification of the N-atom is already favourable with minimal thermal energy (i.e. low temperature), the loss of F requires a moderate energy budget - indicative of a reaction where the mobility of the species on the surface plays a role. At body temperature both of these reactions are favourable and the vast majority of the molecules are seen to have reacted. 
 
X-ray absorption spectroscopy  (XAS) measurements of 5-Fu on Ag also support the notion of a change in the molecular $\pi$-bonding configuration of the molecule (Fig.\ \ref{fig2}f). Compared to the bulk-like sample, 5-Fu on Ag  shows additional features at lower photon energies, in the energy range corresponding to a transition from the N1s core level to an unoccupied $\pi^{\star}$ orbital (indicated by the green area in Fig.\ \ref{fig2}f).  This result suggests that the $\pi$-bonding of an N-atom has been significantly modified, but that a delocalised $\pi$ state  still remains. This view is consistent with the XPS inference of a silver-mediated reaction involving the breaking of an N-H bond. For both the bulk film and 5-Fu on Ag, the $\pi^{\star}$ absorption features are reduced at normal incidence (NI) relative to the measurement at 45$^{\circ}$ incidence, indicating that the molecules have a preference to lie flat on the substrate. For the bulk film, the molecular orientation is broadly consistent with the bulk crystal structure. For the thin film 5-Fu on Ag, the molecular orientation is consistent with STM measurements and the orientation expected for the reacted species (see supplementary material for the bulk crystal structure and STM measurements).

\begin{figure}
\centering
\includegraphics [width=\columnwidth]{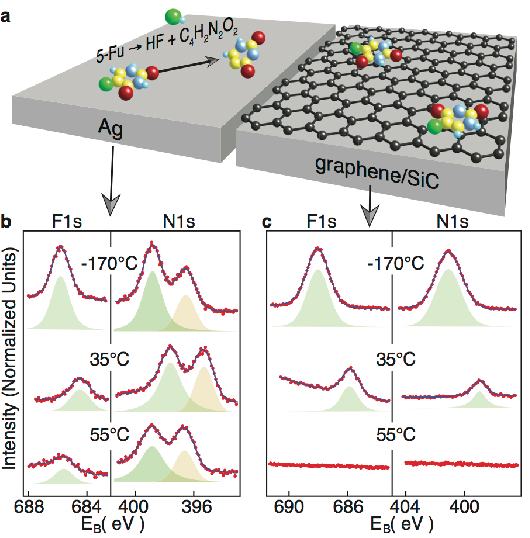}
\caption{\textbf{The interaction of 5-Fu with graphene and silver:} \textbf{a,} Schematic represention of 5-Fu in contact with Ag and graphene surfaces. \textbf{b,}  Temperature dependent F1s and N1s XPS measurements of 5-Fu on Ag. \textbf{c,} The same temperature dependence for 5-Fu on graphene. No reaction is seen, and at $T$=55~$^{\circ}$C the molecules have been completely desorbed, emphasising the weak interaction between 5-Fu and graphene). } \label{fig3}
\end{figure}

The reaction observed on the Ag surface is not seen on graphene. Graphene is chosen because we expect it to be chemically inert, and because it already attracts attention as the ultimately thin protective coating \cite{Prasai:2012,Nilsson:2012, Kirkland:2012}. However, its potential as a coating material in chemotherapeutic drug delivery systems appears to have been entirely overlooked. Graphene is a bio-compatible material \cite{Gurunathan:2013,Fan:2010} with low toxicity. In contrast to many metals, the accidental release of small quantities of a graphene coating into a patient is less concerning than a similar release of alternative coating materials.

A thin film ($\approx$0.7~nm) of  5-Fu on graphene was prepared by thermal evaporation, following the same procedure as above. i.e., the only difference with the experiment of 5-Fu on Ag is that a graphene substrate is used in place of Ag. Temperature dependent XPS data have also been acquired (Fig.\ \ref{fig3}c) and compared with that from 5-Fu on Ag (Fig.\ \ref{fig3}b). Contrary to 5-Fu on Ag, the chemotherapy molecules on graphene do not react at any temperature; instead they are simply desorbed at moderate temperatures. This observation supports the idea of a weak interaction between graphene and 5-Fu since already at $\approx$55~$^{\circ}$C the intact molecules have left the  surface and no  N and F can be detected by XPS.

 \begin{figure*}
\centering
\includegraphics [width=0.8 \textwidth]{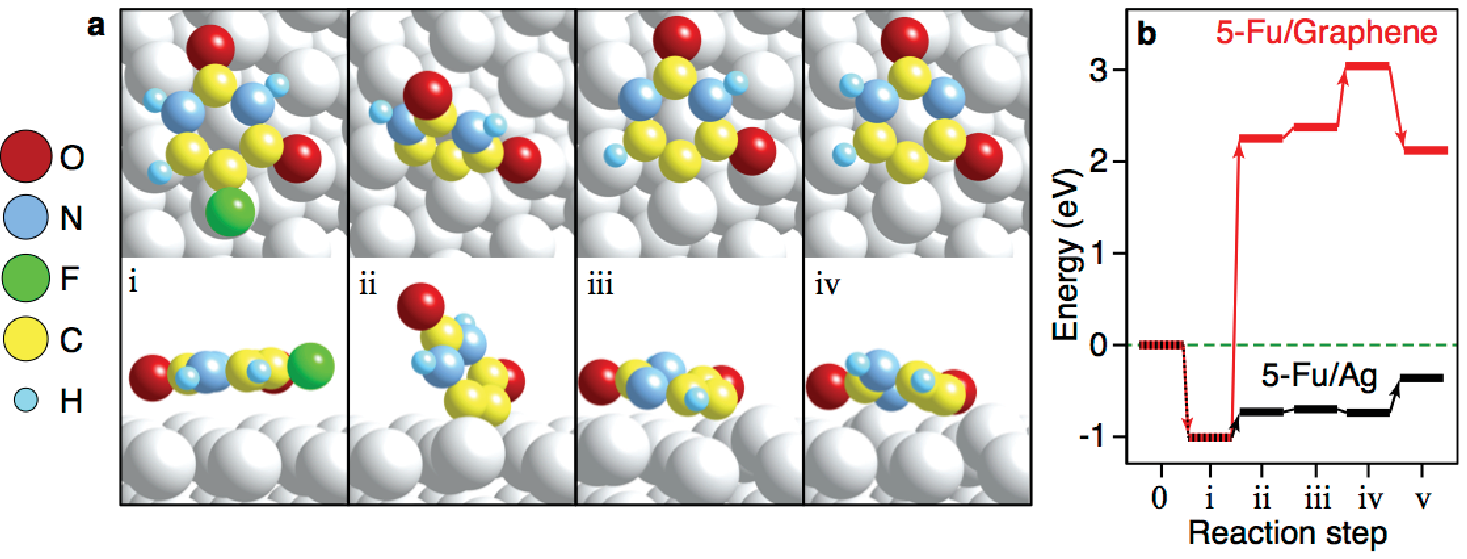}
\caption{\textbf{DFT calculated reaction pathway for 5-Fu on Ag:} \textbf{a} Top view and side view of four of the reaction steps are shown (labelled \textrm{i}-\textrm{iv}). Note; after step \textrm{i}, H- and F-remain bound to the surface, but are not shown in the figure. \textbf{b} The energy diagram for this reaction pathway for both 5-Fu on Ag and 5-Fu on graphene. The initial energy is zero by definition, and the absorption energy of 5-Fu on the two surfaces (step \textrm{i}) is almost identical. However, the energy of the decomposition (step \textrm{ii}) is dramatically different.} \label{dft}
\end{figure*}

DFT calculations have also been performed which support our experimental findings. Fig.\ \ref{dft} shows four of the steps of one possible 5-Fu+Ag reaction pathway. The first step (\textrm{i}) involves the absorption of the molecule onto the Ag(111) surface; it is seen to be weakly physisorbed, and the total energy is lowered by $\approx$1~eV.  In step \textrm{ii}, the C-F bond and a C-H bond are broken, resulting in a small increase in the total energy, but feasible even at low temperature. H and F are liberated from the molecule and stay bound to the Ag surface. In steps \textrm{iii} and \textrm{iv}, the H-atoms which remain attached to the molecule fragment rearrange, with negligible change in the energy. Finally, in step \textrm{v} (not shown), H and F recombine and an HF molecule is desorbed into vacuum. This last step requires a moderate input of energy, and is consistent with the measured temperature dependence. 

To summarize, the molecules react at low temperature and the bonding in the vicinity of one N-atom is modified with the macrocycle remaining intact. Although H and F are already `lost' from the 5-Fu molecule, the HF recombination and liberation from the surface cannot occur without a moderate amount of thermal energy. In agreement with the observation of HF in our residual gas analysis, these calculations support HF, rather than F$_2$, as the favourable product (see the supplementary material for details of the residual gas analysis and the energetics of the F$_2$ formation).



An identical reaction pathway has been calculated for 5-Fu on graphene (see Fig.\ \ref{dft}b). Contrary to the situation for Ag, step \textrm{ii} requires a large energy input ($\approx$3~eV), which is prohibitive. In fact, it is much larger than the absorption energy (step \textrm{i}), hence as the thermal energy is increased, the intact molecule can desorb well before it has the possibility to decompose.

The DFT calculation also confirms that the molecule has a tendency to lie flat on both the Ag and graphene substrate - which is not surprising since molecules with conjugated $\pi$-systems typically show this tendency \cite{Song:2009, Yamane:2007}.  The calculated orientation is consistent with the orientation observed with XAS and STM. Finally, the DFT calculation gives access to the binding energy of the reaction species;  the two inequivalent N-atoms in the reacted molecular fragments have N1s components separated by 1.7~eV (compared to a calculated N1s separation of 0.4~eV for unreacted 5-Fu), this is in good agreement with the experimentally determined value of 2.2~eV.

\section*{Discussion}

Using XPS, XAS and DFT (and supported by supplementary techniques), we observed that a reaction takes place for 5-Fu on silver which does not occur for 5-Fu on graphene. Our experiments have been performed on model surfaces in a well controlled ultrahigh vacuum environment ($P\le 10^{-9}$~mbar). Although the experimental conditions necessarily differ from \textit{in operando} drug delivery (most notably, the lack of ionic solution), it is of concern that an energetically favourable reaction exists between 5-Fu and Ag, (with HF as a product) -- especially when this reaction is readily avoided by choosing a more suitable coating material.


We conclude that silver and argentic alloys might not be a good choice for catheters delivering 5-Fu, since our measurements reveal this chemotherapeutic agent to be catalytically degradeble, with HF produced as a product. Such a reaction occurring in a real catheter would not only compromise the efficacy of the treatment, but also create concern that HF is assimilated into the bloodstream of a patient \cite{Yamashita:2001}. Furthermore, we speculate that the presence of HF in the drug solution accounts, at least partially, for the prevalent degradation of catheters and their coatings. In the case of Ag coatings, this may release a considerable amount of Ag into the bloodstream, also raising significant health concerns \cite{Fung:1996}.  

Since the potential effects of 5-Fu degradation are so significant, it seems surprising that thorough studies of the relevant catheter surface chemistry have not been carried out, and we believe that our study represents an important first step in rectifying this missing knowledge. Furthermore, our study demonstrates more generally that XPS, DFT and other surface chemistry approaches are well suited to study the interaction of chemotherapeutic drugs on relevant surfaces.

Graphene has already been suggested as an external coating for bio-medical applications, and we therefore propose to use graphene for the internal surfaces too. The fabrication of thin carbon-based coatings is technologically realistic \cite{Podila:2013, Bunch:2008, Zhang:2014a}; graphene can even be grown directly on top of silver \cite{Cai:2010,Kiraly:2013}. We therefore believe that graphene coatings will offer a vastly superior alternative to silver-based coatings, and can return a societal benefit within a short timescale. \\

\textbf{Acknowledgements:} The synchrotron access was partially funded under the `CALIPSO' scheme. FM acknowledges travel support from the Norwegian PhD Network on Nanotechnology for Microsystems, sponsored by the Research Council of Norway (contract no. 221860/F40). The calculation power is granted by The Norwegian Metacenter for Computational Science (NOTUR), project numbers: nn9229k and nn4504k. \\

\textbf{Methods:} The `powder' sample was prepared by dissolving 5-Fu powder ($>$99\%, Sigma-Aldrich,  F6627-5G) in de-ionised water, followed by dripping the solution onto a tantalum sample plate and allowing it to dry under atmospheric conditions. After drying, a thick white powder layer was observed.

All other 5-Fu films were made by thermal evaporation under ultra-high vacuum (UHV) conditions. The same 5-Fu powder was placed in a tantalum crucible and heated in-vacuum until it evaporated. The substrate was held in front of the crucible and exposed to the molecular flux. 

Prototype graphene samples were fabricated using standard CVD instrumentation in NTNU's NanoLab. The high quality graphene sample was thermally grown on SiC with standard methods described in our previous work \cite{Mazzola:2013a}.  The Ag sample was a (111) single crystal which had been prepared in-vacuum by cycles of sputtering and annealing.

Photoemission measurements were carried out at beamline D1011 at the MAX IV laboratory (Lund, Sweden). The thermally evaporated films were grown at the beamline and measured without exposure to air.  

Samples were prepared for STM analysis by similar thermal evaporation in UHV.

Calculations were performed using Quantum Espresso \cite{Giannozzi:2009}. We employed the generalized gradient approximation (GGA) method PBE functional \cite{Perdew:1996} with Van der Waals correction \cite{Grimme:2006}. An ultra soft pseudo-potential was used \cite{Rappe:1990}. Satisfactory convergences were obtained with cutoffs of 70~Ry on the plane waves and 700 Ry on the electronic density. These conditions were checked to get a convergence of properties and lattice parameter of Ag bulk of 4.15~{\AA}  (experimental value 4.09~\AA) \cite{Linstrom:2001} We used the $\Gamma$ point for the k-point sampling of the Brillouin zone.

\vspace{-0.15in}

\begin{thebibliography}{39}
\expandafter\ifx\csname natexlab\endcsname\relax\def\natexlab#1{#1}\fi
\expandafter\ifx\csname bibnamefont\endcsname\relax
  \def\bibnamefont#1{#1}\fi
\expandafter\ifx\csname bibfnamefont\endcsname\relax
  \def\bibfnamefont#1{#1}\fi
\expandafter\ifx\csname citenamefont\endcsname\relax
  \def\citenamefont#1{#1}\fi
\expandafter\ifx\csname url\endcsname\relax
  \def\url#1{\texttt{#1}}\fi
\expandafter\ifx\csname urlprefix\endcsname\relax\def\urlprefix{URL }\fi
\providecommand{\bibinfo}[2]{#2}
\providecommand{\eprint}[2][]{\url{#2}}

\bibitem[{\citenamefont{Lu et~al.}(1993)\citenamefont{Lu, Zhang, and
  Diasio}}]{Lu:1993}
\bibinfo{author}{\bibfnamefont{Z.}~\bibnamefont{Lu}},
  \bibinfo{author}{\bibfnamefont{R.}~\bibnamefont{Zhang}}, \bibnamefont{and}
  \bibinfo{author}{\bibfnamefont{R.~B.} \bibnamefont{Diasio}},
  \bibinfo{journal}{Cancer Research} \textbf{\bibinfo{volume}{53}},
  \bibinfo{pages}{5433} (\bibinfo{year}{1993}).

\bibitem[{\citenamefont{Sistigu et~al.}(2014)\citenamefont{Sistigu, Yamazaki,
  Vacchelli, Chaba, Enot, Adam, Vitale, Goubar, Baracco, Remedios
  et~al.}}]{Sistigu:2014}
\bibinfo{author}{\bibfnamefont{A.}~\bibnamefont{Sistigu}},
  \bibinfo{author}{\bibfnamefont{T.}~\bibnamefont{Yamazaki}},
  \bibinfo{author}{\bibfnamefont{E.}~\bibnamefont{Vacchelli}},
  \bibinfo{author}{\bibfnamefont{K.}~\bibnamefont{Chaba}},
  \bibinfo{author}{\bibfnamefont{D.~P.} \bibnamefont{Enot}},
  \bibinfo{author}{\bibfnamefont{J.}~\bibnamefont{Adam}},
  \bibinfo{author}{\bibfnamefont{I.}~\bibnamefont{Vitale}},
  \bibinfo{author}{\bibfnamefont{A.}~\bibnamefont{Goubar}},
  \bibinfo{author}{\bibfnamefont{E.~E.} \bibnamefont{Baracco}},
  \bibinfo{author}{\bibfnamefont{C.}~\bibnamefont{Remedios}},
  \bibnamefont{et~al.}, \bibinfo{journal}{Nat Med}
  \textbf{\bibinfo{volume}{advance online publication}},
  (\bibinfo{year}{2014}), \urlprefix\url{http://dx.doi.org/10.1038/nm.3708}.

\bibitem[{\citenamefont{Longley et~al.}(2003)\citenamefont{Longley, Harkin, and
  Johnston}}]{Longley:2003}
\bibinfo{author}{\bibfnamefont{D.}~\bibnamefont{Longley}},
  \bibinfo{author}{\bibfnamefont{D.}~\bibnamefont{Harkin}}, \bibnamefont{and}
  \bibinfo{author}{\bibfnamefont{P.}~\bibnamefont{Johnston}},
  \bibinfo{journal}{Nature Reviews Cancer} \textbf{\bibinfo{volume}{3}},
  \bibinfo{pages}{330 338} (\bibinfo{year}{2003}).

\bibitem[{\citenamefont{Levine et~al.}(1998)\citenamefont{Levine, Bramwell,
  Pritchard, Norris, Shepherd, Abu-Zahra, Findlay, Warr, Bowman, Myles
  et~al.}}]{Levine:1998}
\bibinfo{author}{\bibfnamefont{M.~N.} \bibnamefont{Levine}},
  \bibinfo{author}{\bibfnamefont{V.~H.} \bibnamefont{Bramwell}},
  \bibinfo{author}{\bibfnamefont{K.~I.} \bibnamefont{Pritchard}},
  \bibinfo{author}{\bibfnamefont{B.~D.} \bibnamefont{Norris}},
  \bibinfo{author}{\bibfnamefont{L.~E.} \bibnamefont{Shepherd}},
  \bibinfo{author}{\bibfnamefont{H.}~\bibnamefont{Abu-Zahra}},
  \bibinfo{author}{\bibfnamefont{B.}~\bibnamefont{Findlay}},
  \bibinfo{author}{\bibfnamefont{D.}~\bibnamefont{Warr}},
  \bibinfo{author}{\bibfnamefont{D.}~\bibnamefont{Bowman}},
  \bibinfo{author}{\bibfnamefont{J.}~\bibnamefont{Myles}},
  \bibnamefont{et~al.}, \bibinfo{journal}{Journal of Clinical Oncology}
  \textbf{\bibinfo{volume}{16}}, \bibinfo{pages}{2651} (\bibinfo{year}{1998}).

\bibitem[{\citenamefont{O'Leary and Bodenham}(2011)}]{OLeary:2011}
\bibinfo{author}{\bibfnamefont{R.}~\bibnamefont{O'Leary}} \bibnamefont{and}
  \bibinfo{author}{\bibfnamefont{A.}~\bibnamefont{Bodenham}},
  \bibinfo{journal}{European Journal of Anaesthesiology 28 (5): 327--8}
  \textbf{\bibinfo{volume}{28}}, \bibinfo{pages}{327} (\bibinfo{year}{2011}).

\bibitem[{\citenamefont{Fackler~Schwalbe
  et~al.}(2004)\citenamefont{Fackler~Schwalbe, Schwalbe, Epple, Becker,
  Pr{\"u}gl, Gassel, Stoffels, and S{\"u}dhoff}}]{Fakler:2004}
\bibinfo{author}{\bibfnamefont{I.}~\bibnamefont{Fackler~Schwalbe}},
  \bibinfo{author}{\bibfnamefont{B.}~\bibnamefont{Schwalbe}},
  \bibinfo{author}{\bibfnamefont{M.}~\bibnamefont{Epple}},
  \bibinfo{author}{\bibfnamefont{A.}~\bibnamefont{Becker}},
  \bibinfo{author}{\bibfnamefont{L.}~\bibnamefont{Pr{\"u}gl}},
  \bibinfo{author}{\bibfnamefont{W.}~\bibnamefont{Gassel}},
  \bibinfo{author}{\bibfnamefont{D.}~\bibnamefont{Stoffels}}, \bibnamefont{and}
  \bibinfo{author}{\bibfnamefont{T.}~\bibnamefont{S{\"u}dhoff}},
  \bibinfo{journal}{Int J Radiat Oncol Biol Phys}
  \textbf{\bibinfo{volume}{89}}, \bibinfo{pages}{547} (\bibinfo{year}{2004}).

\bibitem[{\citenamefont{Ardalan and Flores}(1995)}]{Ardalan:1995}
\bibinfo{author}{\bibfnamefont{B.}~\bibnamefont{Ardalan}} \bibnamefont{and}
  \bibinfo{author}{\bibfnamefont{M.}~\bibnamefont{Flores}},
  \bibinfo{journal}{Cancer} \textbf{\bibinfo{volume}{75}},
  \bibinfo{pages}{2165} (\bibinfo{year}{1995}).

\bibitem[{\citenamefont{O'Grady et~al.}(2011)\citenamefont{O'Grady, Alexander,
  Burns, Dellinger, Garland, Heard, Lipsett, Masur, Mermel, Pearson
  et~al.}}]{OGrady:2011}
\bibinfo{author}{\bibfnamefont{N.~P.} \bibnamefont{O'Grady}},
  \bibinfo{author}{\bibfnamefont{M.}~\bibnamefont{Alexander}},
  \bibinfo{author}{\bibfnamefont{L.~A.} \bibnamefont{Burns}},
  \bibinfo{author}{\bibfnamefont{E.~P.} \bibnamefont{Dellinger}},
  \bibinfo{author}{\bibfnamefont{J.}~\bibnamefont{Garland}},
  \bibinfo{author}{\bibfnamefont{S.~O.} \bibnamefont{Heard}},
  \bibinfo{author}{\bibfnamefont{P.~A.} \bibnamefont{Lipsett}},
  \bibinfo{author}{\bibfnamefont{H.}~\bibnamefont{Masur}},
  \bibinfo{author}{\bibfnamefont{L.~A.} \bibnamefont{Mermel}},
  \bibinfo{author}{\bibfnamefont{M.~L.} \bibnamefont{Pearson}},
  \bibnamefont{et~al.}, \bibinfo{journal}{Clinical Infectious Diseases}
  \textbf{\bibinfo{volume}{52}}, \bibinfo{pages}{e162} (\bibinfo{year}{2011}).

\bibitem[{\citenamefont{Masci et~al.}(2003)\citenamefont{Masci, Magagnoli,
  Zucali, Castagna, Carnaghi, Sarina, Pedicini, Fallini, and
  Santoro}}]{Masci:2003}
\bibinfo{author}{\bibfnamefont{G.}~\bibnamefont{Masci}},
  \bibinfo{author}{\bibfnamefont{M.}~\bibnamefont{Magagnoli}},
  \bibinfo{author}{\bibfnamefont{P.}~\bibnamefont{Zucali}},
  \bibinfo{author}{\bibfnamefont{L.}~\bibnamefont{Castagna}},
  \bibinfo{author}{\bibfnamefont{C.}~\bibnamefont{Carnaghi}},
  \bibinfo{author}{\bibfnamefont{B.}~\bibnamefont{Sarina}},
  \bibinfo{author}{\bibfnamefont{V.}~\bibnamefont{Pedicini}},
  \bibinfo{author}{\bibfnamefont{M.}~\bibnamefont{Fallini}}, \bibnamefont{and}
  \bibinfo{author}{\bibfnamefont{A.}~\bibnamefont{Santoro}},
  \bibinfo{journal}{J. Clin. Oncol.} \textbf{\bibinfo{volume}{21}},
  \bibinfo{pages}{736} (\bibinfo{year}{2003}).

\bibitem[{\citenamefont{Timsit et~al.}(2011)\citenamefont{Timsit, Dubois,
  Minet, Bonadona, Lugosi, Ara-Somohano, Hamidfar-Roy, and
  Schwebel}}]{Timsit:2011}
\bibinfo{author}{\bibfnamefont{J.-F.} \bibnamefont{Timsit}},
  \bibinfo{author}{\bibfnamefont{Y.}~\bibnamefont{Dubois}},
  \bibinfo{author}{\bibfnamefont{C.}~\bibnamefont{Minet}},
  \bibinfo{author}{\bibfnamefont{A.}~\bibnamefont{Bonadona}},
  \bibinfo{author}{\bibfnamefont{M.}~\bibnamefont{Lugosi}},
  \bibinfo{author}{\bibfnamefont{C.}~\bibnamefont{Ara-Somohano}},
  \bibinfo{author}{\bibfnamefont{R.}~\bibnamefont{Hamidfar-Roy}},
  \bibnamefont{and} \bibinfo{author}{\bibfnamefont{C.}~\bibnamefont{Schwebel}},
  \bibinfo{journal}{Annals of Intensive Care} \textbf{\bibinfo{volume}{1}},
  \bibinfo{pages}{1} (\bibinfo{year}{2011}).

\bibitem[{\citenamefont{Armentano et~al.}(2014)\citenamefont{Armentano,
  Arciola, Fortunati, Ferrari, Mattioli, Amoroso, Rizzo, Kenny, Imbriani, and
  Visai}}]{Armentano:2014}
\bibinfo{author}{\bibfnamefont{I.}~\bibnamefont{Armentano}},
  \bibinfo{author}{\bibfnamefont{C.~R.} \bibnamefont{Arciola}},
  \bibinfo{author}{\bibfnamefont{E.}~\bibnamefont{Fortunati}},
  \bibinfo{author}{\bibfnamefont{D.}~\bibnamefont{Ferrari}},
  \bibinfo{author}{\bibfnamefont{S.}~\bibnamefont{Mattioli}},
  \bibinfo{author}{\bibfnamefont{C.~F.} \bibnamefont{Amoroso}},
  \bibinfo{author}{\bibfnamefont{J.}~\bibnamefont{Rizzo}},
  \bibinfo{author}{\bibfnamefont{J.~M.} \bibnamefont{Kenny}},
  \bibinfo{author}{\bibfnamefont{M.}~\bibnamefont{Imbriani}}, \bibnamefont{and}
  \bibinfo{author}{\bibfnamefont{L.}~\bibnamefont{Visai}},
  \bibinfo{journal}{The Scientific World Journal}
  \textbf{\bibinfo{volume}{2014}}, \bibinfo{pages}{18} (\bibinfo{year}{2014}).

\bibitem[{\citenamefont{Suska et~al.}(2010)\citenamefont{Suska, Svensson,
  Johansson, Emanuelsson, Karlholm, Ohrlander, and Thomsen}}]{Suska:2010}
\bibinfo{author}{\bibfnamefont{F.}~\bibnamefont{Suska}},
  \bibinfo{author}{\bibfnamefont{S.}~\bibnamefont{Svensson}},
  \bibinfo{author}{\bibfnamefont{A.}~\bibnamefont{Johansson}},
  \bibinfo{author}{\bibfnamefont{L.}~\bibnamefont{Emanuelsson}},
  \bibinfo{author}{\bibfnamefont{H.}~\bibnamefont{Karlholm}},
  \bibinfo{author}{\bibfnamefont{M.}~\bibnamefont{Ohrlander}},
  \bibnamefont{and} \bibinfo{author}{\bibfnamefont{P.}~\bibnamefont{Thomsen}},
  \bibinfo{journal}{Journal of Biomedical Materials Research Part B: Applied
  Biomaterials} \textbf{\bibinfo{volume}{92B}}, \bibinfo{pages}{86}
  (\bibinfo{year}{2010}), ISSN \bibinfo{issn}{1552-4981}.

\bibitem[{\citenamefont{Harter et~al.}(2002)\citenamefont{Harter, Salwender,
  Bach, Egere, Goldschmidt, and Ho}}]{Harter:2002}
\bibinfo{author}{\bibfnamefont{C.}~\bibnamefont{Harter}},
  \bibinfo{author}{\bibfnamefont{H.}~\bibnamefont{Salwender}},
  \bibinfo{author}{\bibfnamefont{A.}~\bibnamefont{Bach}},
  \bibinfo{author}{\bibfnamefont{G.}~\bibnamefont{Egere}},
  \bibinfo{author}{\bibfnamefont{H.}~\bibnamefont{Goldschmidt}},
  \bibnamefont{and} \bibinfo{author}{\bibfnamefont{A.}~\bibnamefont{Ho}},
  \bibinfo{journal}{Cancer} \textbf{\bibinfo{volume}{94}}, \bibinfo{pages}{245}
  (\bibinfo{year}{2002}).

\bibitem[{\citenamefont{Maki et~al.}(1997)\citenamefont{Maki, Stolz, Wheeler,
  and Mermel}}]{Maki:1997}
\bibinfo{author}{\bibfnamefont{D.}~\bibnamefont{Maki}},
  \bibinfo{author}{\bibfnamefont{S.}~\bibnamefont{Stolz}},
  \bibinfo{author}{\bibfnamefont{S.}~\bibnamefont{Wheeler}}, \bibnamefont{and}
  \bibinfo{author}{\bibfnamefont{L.~A.} \bibnamefont{Mermel}},
  \bibinfo{journal}{Ann Intern Med} \textbf{\bibinfo{volume}{127}},
  \bibinfo{pages}{257} (\bibinfo{year}{1997}).

\bibitem[{\citenamefont{Singh et~al.}(2008)\citenamefont{Singh, Kumar, Kumar,
  and Chacharkar}}]{Singh:2008}
\bibinfo{author}{\bibfnamefont{R.}~\bibnamefont{Singh}},
  \bibinfo{author}{\bibfnamefont{D.}~\bibnamefont{Kumar}},
  \bibinfo{author}{\bibfnamefont{P.}~\bibnamefont{Kumar}}, \bibnamefont{and}
  \bibinfo{author}{\bibfnamefont{M.}~\bibnamefont{Chacharkar}},
  \bibinfo{journal}{J Burn Care Res} \textbf{\bibinfo{volume}{29}},
  \bibinfo{pages}{64} (\bibinfo{year}{2008}).

\bibitem[{\citenamefont{Guggenbichler et~al.}(1999)\citenamefont{Guggenbichler,
  Boswald, Lugauer, and Krall}}]{Guggenbichler:1999}
\bibinfo{author}{\bibfnamefont{J.}~\bibnamefont{Guggenbichler}},
  \bibinfo{author}{\bibfnamefont{M.}~\bibnamefont{Boswald}},
  \bibinfo{author}{\bibfnamefont{S.}~\bibnamefont{Lugauer}}, \bibnamefont{and}
  \bibinfo{author}{\bibfnamefont{T.}~\bibnamefont{Krall}},
  \bibinfo{journal}{Infection} \textbf{\bibinfo{volume}{27}},
  \bibinfo{pages}{S16} (\bibinfo{year}{1999}).

\bibitem[{\citenamefont{Feng et~al.}(2000)\citenamefont{Feng, Wu, Chen, Cui,
  Kim, and Kim}}]{Feng:2000}
\bibinfo{author}{\bibfnamefont{Q.~L.} \bibnamefont{Feng}},
  \bibinfo{author}{\bibfnamefont{J.}~\bibnamefont{Wu}},
  \bibinfo{author}{\bibfnamefont{G.~Q.} \bibnamefont{Chen}},
  \bibinfo{author}{\bibfnamefont{F.~Z.} \bibnamefont{Cui}},
  \bibinfo{author}{\bibfnamefont{T.~N.} \bibnamefont{Kim}}, \bibnamefont{and}
  \bibinfo{author}{\bibfnamefont{T.~O.} \bibnamefont{Kim}},
  \bibinfo{journal}{Journal of Biomedical Materials Research Part A. 2000.
  Volume 52, issue 4. p. 662-668.} \textbf{\bibinfo{volume}{52}},
  \bibinfo{pages}{662} (\bibinfo{year}{2000}).

\bibitem[{\citenamefont{Pavel et~al.}(2006)\citenamefont{Pavel, Cota, Kiefer,
  and C{\^\i}nt{\u a}-P{\^\i}nzaru}}]{Pavel:2006}
\bibinfo{author}{\bibfnamefont{I.}~\bibnamefont{Pavel}},
  \bibinfo{author}{\bibfnamefont{S.}~\bibnamefont{Cota}},
  \bibinfo{author}{\bibfnamefont{W.}~\bibnamefont{Kiefer}}, \bibnamefont{and}
  \bibinfo{author}{\bibfnamefont{S.}~\bibnamefont{C{\^\i}nt{\u
  a}-P{\^\i}nzaru}}, \bibinfo{journal}{Particulate Science and Technology}
  \textbf{\bibinfo{volume}{24}}, \bibinfo{pages}{301} (\bibinfo{year}{2006}).

\bibitem[{\citenamefont{Papageorgiou et~al.}(2012)\citenamefont{Papageorgiou,
  Fischer, Reichert, Diller, Blobner, Klappenberger, Allegretti, Seitsonen, and
  Barth}}]{Papageorgiou:2012}
\bibinfo{author}{\bibfnamefont{A.~C.} \bibnamefont{Papageorgiou}},
  \bibinfo{author}{\bibfnamefont{S.}~\bibnamefont{Fischer}},
  \bibinfo{author}{\bibfnamefont{J.}~\bibnamefont{Reichert}},
  \bibinfo{author}{\bibfnamefont{K.}~\bibnamefont{Diller}},
  \bibinfo{author}{\bibfnamefont{F.}~\bibnamefont{Blobner}},
  \bibinfo{author}{\bibfnamefont{F.}~\bibnamefont{Klappenberger}},
  \bibinfo{author}{\bibfnamefont{F.}~\bibnamefont{Allegretti}},
  \bibinfo{author}{\bibfnamefont{A.~P.} \bibnamefont{Seitsonen}},
  \bibnamefont{and} \bibinfo{author}{\bibfnamefont{J.~V.} \bibnamefont{Barth}},
  \bibinfo{journal}{ACS Nano} \textbf{\bibinfo{volume}{6}},
  \bibinfo{pages}{2477} (\bibinfo{year}{2012}).

\bibitem[{\citenamefont{Prasai et~al.}(2012)\citenamefont{Prasai, Tuberquia,
  Harl, Jennings, and Bolotin}}]{Prasai:2012}
\bibinfo{author}{\bibfnamefont{D.}~\bibnamefont{Prasai}},
  \bibinfo{author}{\bibfnamefont{J.~C.} \bibnamefont{Tuberquia}},
  \bibinfo{author}{\bibfnamefont{R.~R.} \bibnamefont{Harl}},
  \bibinfo{author}{\bibfnamefont{G.~K.} \bibnamefont{Jennings}},
  \bibnamefont{and} \bibinfo{author}{\bibfnamefont{K.~I.}
  \bibnamefont{Bolotin}}, \bibinfo{journal}{ACS Nano}
  \textbf{\bibinfo{volume}{6}}, \bibinfo{pages}{1102} (\bibinfo{year}{2012}).

\bibitem[{\citenamefont{Nilsson et~al.}(2012)\citenamefont{Nilsson, Andersen,
  Balog, L{\ae}gsgaard, Hofmann, Besenbacher, Hammer, Stensgaard, and
  Hornek{\ae}r}}]{Nilsson:2012}
\bibinfo{author}{\bibfnamefont{L.}~\bibnamefont{Nilsson}},
  \bibinfo{author}{\bibfnamefont{M.}~\bibnamefont{Andersen}},
  \bibinfo{author}{\bibfnamefont{R.}~\bibnamefont{Balog}},
  \bibinfo{author}{\bibfnamefont{E.}~\bibnamefont{L{\ae}gsgaard}},
  \bibinfo{author}{\bibfnamefont{P.}~\bibnamefont{Hofmann}},
  \bibinfo{author}{\bibfnamefont{F.}~\bibnamefont{Besenbacher}},
  \bibinfo{author}{\bibfnamefont{B.}~\bibnamefont{Hammer}},
  \bibinfo{author}{\bibfnamefont{I.}~\bibnamefont{Stensgaard}},
  \bibnamefont{and}
  \bibinfo{author}{\bibfnamefont{L.}~\bibnamefont{Hornek{\ae}r}},
  \bibinfo{journal}{ACS Nano} \textbf{\bibinfo{volume}{6}},
  \bibinfo{pages}{10258} (\bibinfo{year}{2012}).

\bibitem[{\citenamefont{Kirkland et~al.}(2012)\citenamefont{Kirkland, Schiller,
  Medhekar, and Birbilis}}]{Kirkland:2012}
\bibinfo{author}{\bibfnamefont{N.~T.} \bibnamefont{Kirkland}},
  \bibinfo{author}{\bibfnamefont{T.}~\bibnamefont{Schiller}},
  \bibinfo{author}{\bibfnamefont{N.}~\bibnamefont{Medhekar}}, \bibnamefont{and}
  \bibinfo{author}{\bibfnamefont{N.}~\bibnamefont{Birbilis}},
  \bibinfo{journal}{Corros. Sci.} \textbf{\bibinfo{volume}{56}},
  \bibinfo{pages}{1} (\bibinfo{year}{2012}).

\bibitem[{\citenamefont{Gurunathan et~al.}(2013)\citenamefont{Gurunathan, Han,
  Eppakayala, Dayem, Kwon, and Kim}}]{Gurunathan:2013}
\bibinfo{author}{\bibfnamefont{S.}~\bibnamefont{Gurunathan}},
  \bibinfo{author}{\bibfnamefont{J.~W.} \bibnamefont{Han}},
  \bibinfo{author}{\bibfnamefont{V.}~\bibnamefont{Eppakayala}},
  \bibinfo{author}{\bibfnamefont{A.~A.} \bibnamefont{Dayem}},
  \bibinfo{author}{\bibfnamefont{N.~D.} \bibnamefont{Kwon}}, \bibnamefont{and}
  \bibinfo{author}{\bibfnamefont{J.~H.} \bibnamefont{Kim}},
  \bibinfo{journal}{Nanoscale Research Letters} \textbf{\bibinfo{volume}{8}}
  (\bibinfo{year}{2013}).

\bibitem[{\citenamefont{Fan et~al.}(2010)\citenamefont{Fan, Wang, Zhao, Li,
  Shi, Ge, and Jin}}]{Fan:2010}
\bibinfo{author}{\bibfnamefont{H.}~\bibnamefont{Fan}},
  \bibinfo{author}{\bibfnamefont{L.}~\bibnamefont{Wang}},
  \bibinfo{author}{\bibfnamefont{K.}~\bibnamefont{Zhao}},
  \bibinfo{author}{\bibfnamefont{N.}~\bibnamefont{Li}},
  \bibinfo{author}{\bibfnamefont{Z.}~\bibnamefont{Shi}},
  \bibinfo{author}{\bibfnamefont{Z.}~\bibnamefont{Ge}}, \bibnamefont{and}
  \bibinfo{author}{\bibfnamefont{Z.}~\bibnamefont{Jin}},
  \bibinfo{journal}{Biomacromolecules} \textbf{\bibinfo{volume}{11}},
  \bibinfo{pages}{2345} (\bibinfo{year}{2010}).

\bibitem[{\citenamefont{Song et~al.}(2009)\citenamefont{Song, Wells, Handrup,
  Li, Bao, Schulte, Ahola-Tuomi, Mayor, Swarbrick, Perkins et~al.}}]{Song:2009}
\bibinfo{author}{\bibfnamefont{F.}~\bibnamefont{Song}},
  \bibinfo{author}{\bibfnamefont{J.~W.} \bibnamefont{Wells}},
  \bibinfo{author}{\bibfnamefont{K.}~\bibnamefont{Handrup}},
  \bibinfo{author}{\bibfnamefont{Z.~S.} \bibnamefont{Li}},
  \bibinfo{author}{\bibfnamefont{S.~N.} \bibnamefont{Bao}},
  \bibinfo{author}{\bibfnamefont{K.}~\bibnamefont{Schulte}},
  \bibinfo{author}{\bibfnamefont{M.}~\bibnamefont{Ahola-Tuomi}},
  \bibinfo{author}{\bibfnamefont{L.~C.} \bibnamefont{Mayor}},
  \bibinfo{author}{\bibfnamefont{J.~C.} \bibnamefont{Swarbrick}},
  \bibinfo{author}{\bibfnamefont{E.~W.} \bibnamefont{Perkins}},
  \bibnamefont{et~al.}, \bibinfo{journal}{Nature Nanotechnology}
  \textbf{\bibinfo{volume}{4}}, \bibinfo{pages}{373} (\bibinfo{year}{2009}).

\bibitem[{\citenamefont{Yamane et~al.}(2007)\citenamefont{Yamane, Yoshimura,
  Kawabe, Sumii, Kanai, Ouchi, Ueno, and Seki}}]{Yamane:2007}
\bibinfo{author}{\bibfnamefont{H.}~\bibnamefont{Yamane}},
  \bibinfo{author}{\bibfnamefont{D.}~\bibnamefont{Yoshimura}},
  \bibinfo{author}{\bibfnamefont{E.}~\bibnamefont{Kawabe}},
  \bibinfo{author}{\bibfnamefont{R.}~\bibnamefont{Sumii}},
  \bibinfo{author}{\bibfnamefont{K.}~\bibnamefont{Kanai}},
  \bibinfo{author}{\bibfnamefont{Y.}~\bibnamefont{Ouchi}},
  \bibinfo{author}{\bibfnamefont{N.}~\bibnamefont{Ueno}}, \bibnamefont{and}
  \bibinfo{author}{\bibfnamefont{K.}~\bibnamefont{Seki}},
  \bibinfo{journal}{Physical Review B} \textbf{\bibinfo{volume}{76}},
  \bibinfo{eid}{165436} (pages~\bibinfo{numpages}{10}) (\bibinfo{year}{2007}).

\bibitem[{\citenamefont{Yamashita et~al.}(2001)\citenamefont{Yamashita, Suzuki,
  Hirai, and Kajigaya}}]{Yamashita:2001}
\bibinfo{author}{\bibfnamefont{M.}~\bibnamefont{Yamashita}},
  \bibinfo{author}{\bibfnamefont{M.}~\bibnamefont{Suzuki}},
  \bibinfo{author}{\bibfnamefont{H.}~\bibnamefont{Hirai}}, \bibnamefont{and}
  \bibinfo{author}{\bibfnamefont{H.}~\bibnamefont{Kajigaya}},
  \bibinfo{journal}{Crit. Care Med.} \textbf{\bibinfo{volume}{29}}
  (\bibinfo{year}{2001}).

\bibitem[{\citenamefont{Fung and Bowen}(34)}]{Fung:1996}
\bibinfo{author}{\bibfnamefont{M.}~\bibnamefont{Fung}} \bibnamefont{and}
  \bibinfo{author}{\bibfnamefont{D.}~\bibnamefont{Bowen}},
  \bibinfo{journal}{Journal of Toxicology Clinical Toxicology}
  \textbf{\bibinfo{volume}{1}} (\bibinfo{year}{34}).

\bibitem[{\citenamefont{Podila et~al.}(2013)\citenamefont{Podila, Moore,
  Alexis, and Rao}}]{Podila:2013}
\bibinfo{author}{\bibfnamefont{R.}~\bibnamefont{Podila}},
  \bibinfo{author}{\bibfnamefont{T.}~\bibnamefont{Moore}},
  \bibinfo{author}{\bibfnamefont{F.}~\bibnamefont{Alexis}}, \bibnamefont{and}
  \bibinfo{author}{\bibfnamefont{A.~M.} \bibnamefont{Rao}},
  \bibinfo{journal}{R. Soc. Chem. Adv.} \textbf{\bibinfo{volume}{3}},
  \bibinfo{pages}{1660} (\bibinfo{year}{2013}).

\bibitem[{\citenamefont{Bunch et~al.}(2008)\citenamefont{Bunch, Verbridge,
  Alden, van~der Zande, Parpia, Craighead, and McEuen}}]{Bunch:2008}
\bibinfo{author}{\bibfnamefont{J.~S.} \bibnamefont{Bunch}},
  \bibinfo{author}{\bibfnamefont{S.~S.} \bibnamefont{Verbridge}},
  \bibinfo{author}{\bibfnamefont{J.~S.} \bibnamefont{Alden}},
  \bibinfo{author}{\bibfnamefont{A.~M.} \bibnamefont{van~der Zande}},
  \bibinfo{author}{\bibfnamefont{J.~M.} \bibnamefont{Parpia}},
  \bibinfo{author}{\bibfnamefont{H.~G.} \bibnamefont{Craighead}},
  \bibnamefont{and} \bibinfo{author}{\bibfnamefont{P.~L.}
  \bibnamefont{McEuen}}, \bibinfo{journal}{Nano Letters}
  \textbf{\bibinfo{volume}{8}}, \bibinfo{pages}{2458} (\bibinfo{year}{2008}).

\bibitem[{\citenamefont{Zhang et~al.}(2014)\citenamefont{Zhang, Lee, McNear,
  Chung, Lee, Lee, Crist, Ratliff, Zhong, Chen et~al.}}]{Zhang:2014a}
\bibinfo{author}{\bibfnamefont{W.}~\bibnamefont{Zhang}},
  \bibinfo{author}{\bibfnamefont{S.}~\bibnamefont{Lee}},
  \bibinfo{author}{\bibfnamefont{K.}~\bibnamefont{McNear}},
  \bibinfo{author}{\bibfnamefont{T.}~\bibnamefont{Chung}},
  \bibinfo{author}{\bibfnamefont{S.}~\bibnamefont{Lee}},
  \bibinfo{author}{\bibfnamefont{K.}~\bibnamefont{Lee}},
  \bibinfo{author}{\bibfnamefont{S.~A.} \bibnamefont{Crist}},
  \bibinfo{author}{\bibfnamefont{T.}~\bibnamefont{Ratliff}},
  \bibinfo{author}{\bibfnamefont{Z.}~\bibnamefont{Zhong}},
  \bibinfo{author}{\bibfnamefont{Y.}~\bibnamefont{Chen}}, \bibnamefont{et~al.},
  \bibinfo{journal}{Scientific Reports} \textbf{\bibinfo{volume}{4}}
  (\bibinfo{year}{2014}).

\bibitem[{\citenamefont{Cai et~al.}(2010)\citenamefont{Cai, Ruffieux, Jaafar,
  Bieri, Braun, Blankenburg, Muoth, Seitsonen, Saleh, Feng et~al.}}]{Cai:2010}
\bibinfo{author}{\bibfnamefont{J.}~\bibnamefont{Cai}},
  \bibinfo{author}{\bibfnamefont{P.}~\bibnamefont{Ruffieux}},
  \bibinfo{author}{\bibfnamefont{R.}~\bibnamefont{Jaafar}},
  \bibinfo{author}{\bibfnamefont{M.}~\bibnamefont{Bieri}},
  \bibinfo{author}{\bibfnamefont{T.}~\bibnamefont{Braun}},
  \bibinfo{author}{\bibfnamefont{S.}~\bibnamefont{Blankenburg}},
  \bibinfo{author}{\bibfnamefont{M.}~\bibnamefont{Muoth}},
  \bibinfo{author}{\bibfnamefont{A.~P.} \bibnamefont{Seitsonen}},
  \bibinfo{author}{\bibfnamefont{M.}~\bibnamefont{Saleh}},
  \bibinfo{author}{\bibfnamefont{X.}~\bibnamefont{Feng}}, \bibnamefont{et~al.},
  \bibinfo{journal}{Nature} \textbf{\bibinfo{volume}{466}},
  \bibinfo{pages}{470} (\bibinfo{year}{2010}).

\bibitem[{\citenamefont{Kiraly et~al.}(2013)\citenamefont{Kiraly, Iski, Mannix,
  Fisher, Hersam, and Guisinger}}]{Kiraly:2013}
\bibinfo{author}{\bibfnamefont{B.}~\bibnamefont{Kiraly}},
  \bibinfo{author}{\bibfnamefont{E.~V.} \bibnamefont{Iski}},
  \bibinfo{author}{\bibfnamefont{A.~J.} \bibnamefont{Mannix}},
  \bibinfo{author}{\bibfnamefont{B.~L.} \bibnamefont{Fisher}},
  \bibinfo{author}{\bibfnamefont{M.~C.} \bibnamefont{Hersam}},
  \bibnamefont{and} \bibinfo{author}{\bibfnamefont{N.~P.}
  \bibnamefont{Guisinger}}, \bibinfo{journal}{Nat Commun}
  \textbf{\bibinfo{volume}{4}} (\bibinfo{year}{2013}).

\bibitem[{\citenamefont{Mazzola et~al.}(2013)\citenamefont{Mazzola, Wells,
  Yakimova, Ulstrup, Miwa, Balog, Bianchi, Leandersson, Adell, Hofmann
  et~al.}}]{Mazzola:2013a}
\bibinfo{author}{\bibfnamefont{F.}~\bibnamefont{Mazzola}},
  \bibinfo{author}{\bibfnamefont{J.~W.} \bibnamefont{Wells}},
  \bibinfo{author}{\bibfnamefont{R.}~\bibnamefont{Yakimova}},
  \bibinfo{author}{\bibfnamefont{S.}~\bibnamefont{Ulstrup}},
  \bibinfo{author}{\bibfnamefont{J.~A.} \bibnamefont{Miwa}},
  \bibinfo{author}{\bibfnamefont{R.}~\bibnamefont{Balog}},
  \bibinfo{author}{\bibfnamefont{M.}~\bibnamefont{Bianchi}},
  \bibinfo{author}{\bibfnamefont{M.}~\bibnamefont{Leandersson}},
  \bibinfo{author}{\bibfnamefont{J.}~\bibnamefont{Adell}},
  \bibinfo{author}{\bibfnamefont{P.}~\bibnamefont{Hofmann}},
  \bibnamefont{et~al.}, \bibinfo{journal}{Phys. Rev. Lett.}
  \textbf{\bibinfo{volume}{111}}, \bibinfo{pages}{216806}
  (\bibinfo{year}{2013}).

\bibitem[{\citenamefont{Giannozzi et~al.}(2009)\citenamefont{Giannozzi, Baroni,
  Bonini, Calandra, Car, Cavazzoni, Ceresoli, Chiarotti, Cococcioni, Dabo
  et~al.}}]{Giannozzi:2009}
\bibinfo{author}{\bibfnamefont{P.}~\bibnamefont{Giannozzi}},
  \bibinfo{author}{\bibfnamefont{S.}~\bibnamefont{Baroni}},
  \bibinfo{author}{\bibfnamefont{N.}~\bibnamefont{Bonini}},
  \bibinfo{author}{\bibfnamefont{M.}~\bibnamefont{Calandra}},
  \bibinfo{author}{\bibfnamefont{R.}~\bibnamefont{Car}},
  \bibinfo{author}{\bibfnamefont{C.}~\bibnamefont{Cavazzoni}},
  \bibinfo{author}{\bibfnamefont{D.}~\bibnamefont{Ceresoli}},
  \bibinfo{author}{\bibfnamefont{G.~L.} \bibnamefont{Chiarotti}},
  \bibinfo{author}{\bibfnamefont{M.}~\bibnamefont{Cococcioni}},
  \bibinfo{author}{\bibfnamefont{I.}~\bibnamefont{Dabo}}, \bibnamefont{et~al.},
  \bibinfo{journal}{Journal of Physics: Condensed Matter}
  \textbf{\bibinfo{volume}{21}}, \bibinfo{pages}{395502}
  (\bibinfo{year}{2009}).

\bibitem[{\citenamefont{Perdew et~al.}(1996)\citenamefont{Perdew, Burke, and
  Ernzerhof}}]{Perdew:1996}
\bibinfo{author}{\bibfnamefont{J.~P.} \bibnamefont{Perdew}},
  \bibinfo{author}{\bibfnamefont{K.}~\bibnamefont{Burke}}, \bibnamefont{and}
  \bibinfo{author}{\bibfnamefont{M.}~\bibnamefont{Ernzerhof}},
  \bibinfo{journal}{Phys. Rev. Lett.} \textbf{\bibinfo{volume}{77}},
  \bibinfo{pages}{3865} (\bibinfo{year}{1996}).

\bibitem[{\citenamefont{Grimme}(2006)}]{Grimme:2006}
\bibinfo{author}{\bibfnamefont{S.}~\bibnamefont{Grimme}},
  \bibinfo{journal}{Journal of Computational Chemistry}
  \textbf{\bibinfo{volume}{27}}, \bibinfo{pages}{1787} (\bibinfo{year}{2006}).

\bibitem[{\citenamefont{Rappe et~al.}(1990)\citenamefont{Rappe, Rabe, Kaxiras,
  and Joannopoulos}}]{Rappe:1990}
\bibinfo{author}{\bibfnamefont{A.~M.} \bibnamefont{Rappe}},
  \bibinfo{author}{\bibfnamefont{K.~M.} \bibnamefont{Rabe}},
  \bibinfo{author}{\bibfnamefont{E.}~\bibnamefont{Kaxiras}}, \bibnamefont{and}
  \bibinfo{author}{\bibfnamefont{J.~D.} \bibnamefont{Joannopoulos}},
  \bibinfo{journal}{Phys. Rev. B} \textbf{\bibinfo{volume}{41}},
  \bibinfo{pages}{1227} (\bibinfo{year}{1990}).

\bibitem[{\citenamefont{Linstrom and Mallard}(2001)}]{Linstrom:2001}
\bibinfo{editor}{\bibfnamefont{P.}~\bibnamefont{Linstrom}} \bibnamefont{and}
  \bibinfo{editor}{\bibfnamefont{W.}~\bibnamefont{Mallard}}, eds.,
  \emph{\bibinfo{title}{NIST Chemistry WebBook, NIST Standard Reference
  Database Number 69}} (\bibinfo{publisher}{National Institute of Standards and
  Technology, Gaithersburg MD, 20899}, \bibinfo{year}{2001}),
  \urlprefix\url{http://webbook.nist.gov}.

\end{thebibliography}

\end{document}